\documentclass[
 reprint,
superscriptaddress,
 amsmath,amssymb,
 aps,%linenumbers,
superscriptaddress,
 amsmath,amssymb,
 aps,longbibliography,
floatfix,
]{revtex4-1}

\usepackage{graphicx}% Include figure files
\usepackage{dcolumn}% Align table columns on decimal point
\usepackage{bm}% bold math
\usepackage{amsmath,amssymb,amstext,amsthm}
\usepackage{braket}
\usepackage{bbold}
\usepackage{bbm}
\usepackage{xcolor}
\usepackage{ gensymb }
\usepackage{soul}
\setstcolor{red}
\newcommand{\invdeg}{(\degree)^{-1}}
\newcommand{\coMP}{\text{M}}

\begin{document}

\preprint{APS/123-QED}
%\linenumbers
\setlength{\abovedisplayskip}{1pt}
\title{Characterizing the circularly-oriented macular pigment using spatiotemporal sensitivity to structured light entoptic phenomena}

\author{D. A. Pushin}
\email{dmitry.pushin@uwaterloo.ca}
\affiliation{Institute for Quantum Computing, University of Waterloo,  Waterloo, ON, Canada, N2L3G1}
\affiliation{Department of Physics, University of Waterloo, Waterloo, ON, Canada, N2L3G1}
\affiliation{Centre for Eye and Vision Research, 17W Hong Kong Science Park, Hong Kong}
\affiliation{Incoherent Vision Inc., Wellesley, ON, Canada, N0B2T0}

\author{D. V. Garrad}
\affiliation{Department of Physics, University of Waterloo, Waterloo, ON, Canada, N2L3G1}

\author{C. Kapahi} 
\affiliation{Institute for Quantum Computing, University of Waterloo,  Waterloo, ON, Canada, N2L3G1}
\affiliation{Department of Physics, University of Waterloo, Waterloo, ON, Canada, N2L3G1}
\author{A. E. Silva} 
\affiliation{School of Optometry and Vision Science, University of Waterloo, Waterloo, ON, Canada, N2L3G1}
\author{P. Chahal}
\affiliation{Department of Physics, University at Buffalo, State University of New York, Buffalo, New York 14260, USA}
\author{D. G. Cory}
\affiliation{Institute for Quantum Computing, University of Waterloo,  Waterloo, ON, Canada, N2L3G1}
\affiliation{Department of Chemistry, University of Waterloo, Waterloo, ON, Canada, N2L3G1}
\author{M. Kulmaganbetov}
\affiliation{Centre for Eye and Vision Research, 17W Hong Kong Science Park, Hong Kong}

\author{I. Salehi} 
\affiliation{School of Optometry and Vision Science, University of Waterloo, Waterloo, ON, Canada, N2L3G1}
\author{M. Mungalsingh}
\affiliation{School of Optometry and Vision Science, University of Waterloo, Waterloo, ON, Canada, N2L3G1}
\author{T. Singh}
\affiliation{Centre for Eye and Vision Research, 17W Hong Kong Science Park, Hong Kong}
\author{B. Thompson}
\affiliation{Centre for Eye and Vision Research, 17W Hong Kong Science Park, Hong Kong}
\affiliation{School of Optometry and Vision Science, University of Waterloo, Waterloo, ON, Canada, N2L3G1}
\author{D. Sarenac}
\email{dusansar@buffalo.edu}
\affiliation{Centre for Eye and Vision Research, 17W Hong Kong Science Park, Hong Kong}
\affiliation{Incoherent Vision Inc., Wellesley, ON, Canada, N0B2T0}
\affiliation{Department of Physics, University at Buffalo, State University of New York, Buffalo, New York 14260, USA}

\date{\today}% It is always \today, today,
             %  but any date may be explicitly specified

%\begin{abstract}
%\end{abstract}

\pacs{Valid PACS appear here}

%\section{\label{sec:level1}Introduction}

\begin{abstract}

%The macular pigment (MP) in the radially-oriented Henle fibers that surround the foveola provides signatures of eye diseases and central field dysfunctions; most notably, age-related macular degeneration (AMD), a globally leading cause of irreversible blindness. It also enables the ability to perceive the orientation of polarized blue light through an entoptic phenomena known as the Haidinger's brush. 
The macular pigment (MP) in the radially-oriented Henle fibers that surround the foveola enables the ability to perceive the orientation of polarized blue light through an entoptic phenomena known as the Haidinger's brush. The MP has been linked to eye diseases and central field dysfunctions, most notably age-related macular degeneration (AMD), a globally leading cause of irreversible blindness. Recent integration of structured light techniques into vision science has allowed for the development of more selective and versatile entoptic probes of eye health that provide interpretable thresholds. For example, it enabled the use of variable spatial frequencies and arbitrary obstructions in the presented stimuli. Additionally, it expanded the $\approx2\degree$ retinal eccentricity extent of the Haidinger's brush to $\approx5\degree$ for a similar class of fringe-based stimuli. In this work, we develop a spatiotemporal sensitivity model that maps perceptual thresholds of entoptic phenomenon to the underlying MP structure that supports its perception. We therefore selectively characterize the circularly-oriented macular pigment optical density (coMPOD) rather than total MPOD as typically measured, providing an additional quantification of macular health. %The model predicts the observed increase in size of entoptic phenomena with higher spatial frequencies and provides a tool for reconstructing the spatial profiles of individuals' coMPOD. 
A study was performed where the retinal eccentricity thresholds were measured for five structured light stimuli with unique spatiotemporal frequencies. The results from  fifteen healthy young participants indicate that the coMPOD is inversely proportional to retinal eccentricity in the range of $1.5\degree$ to $5.5\degree$. Good agreement between the model and the collected data is found with a Pearson $\chi^2$ fit statistic of 0.06. The presented techniques can be applied in novel early diagnostic tests for a variety of diseases related to macular degeneration. Specific use-cases include diagnosing AMD, macular telangiectasia, and pathological myopia which are diseases marked by macular deformation.

\end{abstract}
\maketitle

\section{Introduction}

Xanthophylls lutein, zeaxanthin, and meso-zeaxanthin constitute the macular pigment (MP) in the retina that provides protection from the accumulation of photochemical damage~\cite{boulton2001retinal}. A portion of MP is in a layer of radially-oriented Henle fibers, where the MP is on average oriented perpendicularly to the Henle fibres~\cite{grudzinski2017localization}. We will refer to this subset of MP as the circularly-oriented macular pigment (coMP). The effects arising from this arrangement as well as MP's inherent dichroism and absorption spectrum, which peaks with wavelengths around 450~nm, combine to act as a weak radial polarization filter in the eye for blue light~\cite{horvath2004polarized}. Thus a faint entoptic phenomenon known as the Haidinger's brush appears within $\approx2\degree$ retinal eccentricity when uniformly polarized blue light is viewed~\cite{haidinger1844ueber,misson2020polarization,o2021seeing,mottes2022haidinger}. The size of polarization-related entoptic phenomena can be enhanced to $\approx 5\degree$ retinal eccentricity through the use of structured light~\cite{bliokh2023roadmap,kapahi2024measuring}. Further, while Haidinger's brush traditionally contains 2 azimuthal fringes, structured light based stimuli can be constructed to produce percepts with arbitrary numbers of fringes and occlusions to probe polarized light perception with greater detail~\cite{sarenac2020direct,sarenac2022human,pushin2023structured,pushin2024psychophysical}.

\begin{figure*}
    \centering
    \includegraphics[width=\linewidth]{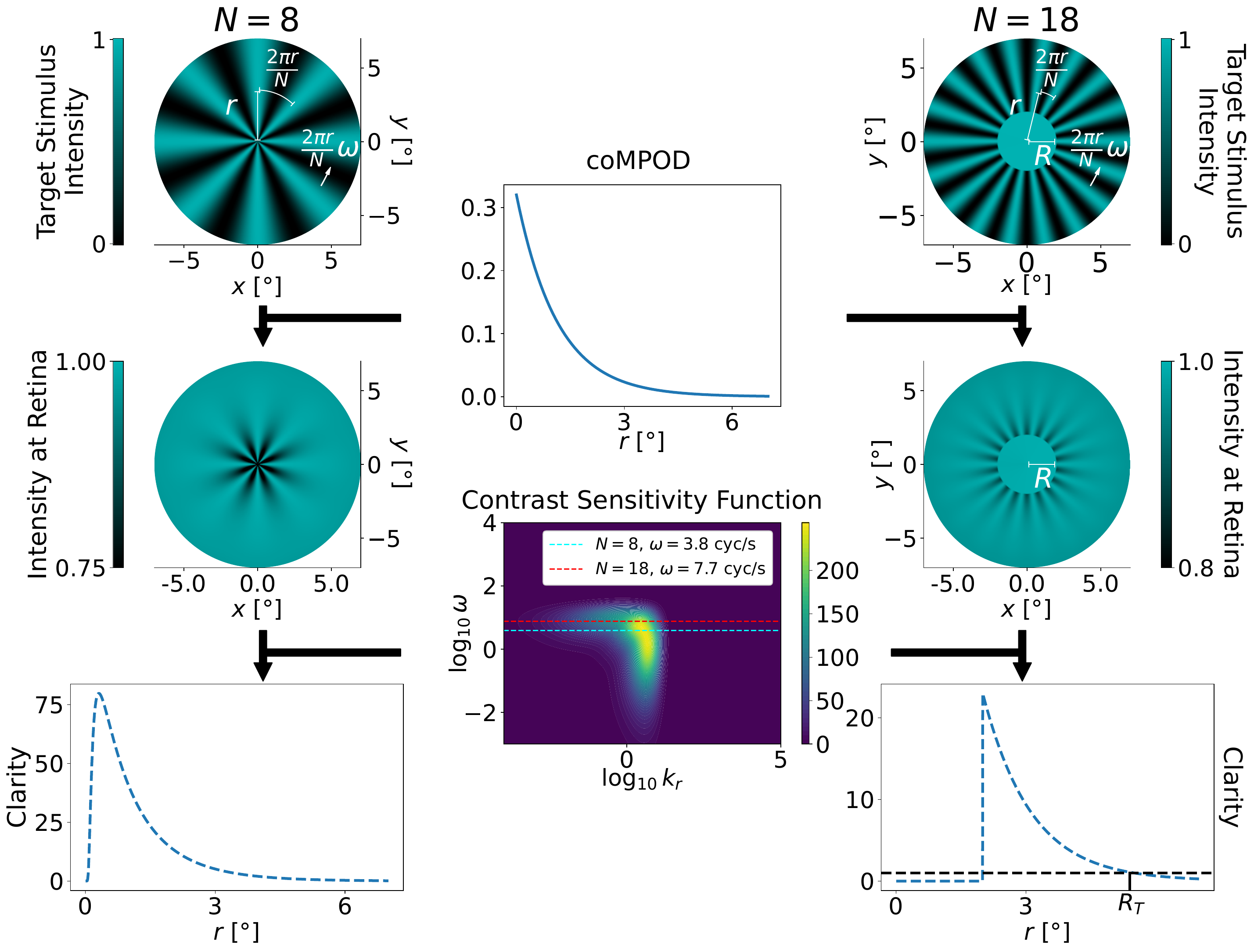}
    \caption{In this work we characterize the observer's circularly-oriented macular pigment (coMP) profile in the range of $1.5\degree$ to $5.5\degree$ retinal eccentricity by measuring the perceptual thresholds of a set of stimuli with unique spatiotemporal frequencies. The target stimulus consists of a central obstruction with radius $R$ and $N$ azimuthal fringes rotating with temporal frequency $\omega$. The top row shows two examples of an intensity pattern where an ideal radial polarizer is considered. The efficiency of the radial polarizer in the human eye is determined by the amount of coMP, which decays with retinal eccentricity. It follows that the contrast of the pattern that a person perceives is highest near the center. To determine a person's ability to perceive a moving pattern we account for the viewer's contrast sensitivity, given by the $M$-scaled contrast sensitivity function ($\text{CSF}_\text{M}[r,k_r,\omega]$) which is equal to the inverse of contrast threshold required for perceiving a particular spatial frequency ($k_r$) and temporal frequency ($\omega$) at retinal eccentricity ($r$). Therefore, the spatially dependent clarity ($C[r]$) of the observed patterns is obtained by multiplying the azimuthal contrast of the stimulus at each radius $r$ by its corresponding value in the $\text{CSF}_\text{M}$. The partially blocked stimulus is considered resolvable for all obstructions with radii $R\leq R_T$ where $C(R_T)=1$. }
    \label{fig:fig_1}
\end{figure*}

Low concentrations of MP have been linked to a higher risk of developing age-related macular
degeneration (AMD)~\cite{kar2020local,landrum1996macular,ozyurt2017comparison,scripsema2015lutein,wu2015intakes,bernstein2002resonance}. AMD is a globally leading cause of irreversible blindness~\cite{thomas2021age} which is often left undiagnosed until after the disease is substantially progressed and visual impairment has begun~\cite{cervantes2008lack}. It has been shown that when detected in their early (often asymptomatic) stages the disease can be treated to slow its progression or prevent further visual impairment~\cite{lombardo2022challenges,nagai2020macular}. However, early detection is challenging because current gold-standard ocular imaging methods such as optical coherence tomography (OCT) are typically limited to detecting drusen and other major structural defects that occur in more advanced AMD~\cite{gerson2017early,neely2017prevalence}. Techniques relying on the perception of uniformly polarized light through the Haidinger's brush to detect the earliest signs of AMD have historically found limited success due to low visibility and lack of interpretable thresholds~\cite{temple2015perceiving,muller2016perception,misson2017spectral,temple2019haidinger}. 

Here, we present a novel method to quantify macular health through the selective characterization of the circularly-oriented macular pigment optical density (coMPOD). The method relies on the perceptual threshold measurements of several structured light stimuli with different spatiotemporal frequencies.  A study was performed with healthy young participants and found that the coMPOD is inversely proportional to retinal eccentricity in the region $1.5\degree$ to $5.5\degree$. The presented techniques, being selectively sensitive to coMP rather than total MP, provide a unique characterization tool for macular health. Potential applications include the detection of early symptoms of a variety of diseases related to macular deterioration, such as pathological myopia~\cite{ueta2020pathologic}, macular telangiectasia~\cite{issa2013macular}, and AMD~\cite{thomas2021age}. 

\section{Methods}
Twenty-four participants with healthy retinas were recruited for the study. The study protocols conformed with the Declaration of Helsinki and each participant provided informed consent prior to the study. Ocular examinations, which included assessments of unaided logMAR distance visual acuity, binocular vision, subjective refraction, ocular motility, slit-lamp biomicroscopy, indirect opthalmoscopy, and colour fundus photography, were performed on each participant during screening. Participants with a previous diagnosis of ocular, systemic, or neurological disorders were excluded. All research procedures received approval from the University of Waterloo Office of Research Ethics.

Participants were asked to perform a psychophysical rotation direction discrimination task. The rotation is necessary because a static polarization-related entoptic phenomena disappears after a few seconds due to visual adaptation~\cite{horvath2004polarized}. The setup that prepares a visual stimulus inducing an entoptic phenomena with variable azimuthal fringes and arbitrary occlusions is described in Ref.~\cite{kapahi2024measuring}. In this study, stimuli with $N=3,8,11,13,\text{ and }18$ azimuthal fringes were generated. Aligned to the center of the stimuli a 50 µm pinhole was illuminated by a red laser to generate an approximately Gaussian guide light with a maximum retinal eccentricity of 0.5$\degree$ that participants were directed to fixate on. Each trial began with the presentation of a static stimuli to ensure visibility, then upon the participant's signal, the rotating (either clockwise or counter-clockwise) stimulus was presented for 500 ms. After the rotating stimulus was presented, the participant indicated the direction of rotation. 

\begin{figure}
    \centering
    \includegraphics[width=\linewidth]{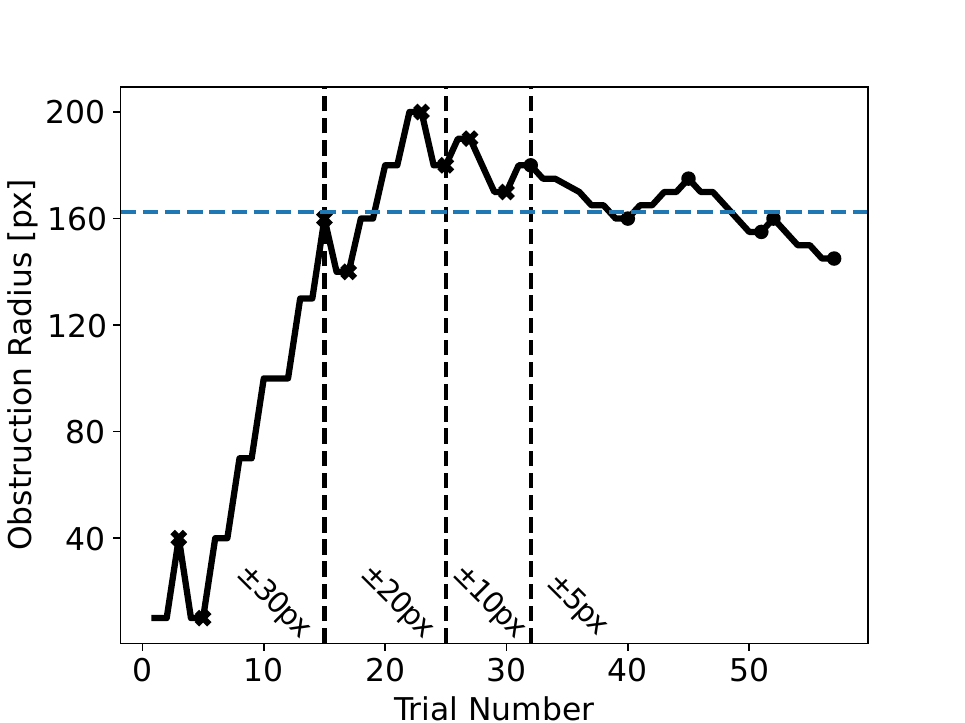}
    \caption{Example of a 2-up/1-down staircase protocol. The reversal points are marked with black points and crosses for clarity, and the change in the obstruction size with each trial is noted in each region. 10 pixels on the Spatial Light Modulator is around $0.45\degree$ retinal eccentricity. After 14 staircase reversals or 90 trials, the arithmetic mean of the final six reversals (marked by black points) is taken to be the participant's threshold radius (blue dashed line) for the given stimulus.}
    \label{fig:fig_4}
\end{figure}

To measure the retinal eccentricity thresholds a circular obstruction with variable size was centered on the fixation point (see Fig.~\ref{fig:fig_1}). The preparation of these structured light states was accomplished via a Spatial Light Modulator whose pixels can be individually addressed to set arbitrary spatial polarization states~\cite{pushin2023structured}. Prior to taking measurements, the participants performed a familiarization task. In this task, the initial obstruction radius was set to roughly $0.45\degree$ and the stimulus was shown for several seconds a total of 10 times. The task was repeated until each participant achieved at least 70\% discrimination accuracy. Following the initial familiarization, the radius of the obstruction, $R$, was changed according to the 2-up/1-down staircase method described in Ref.~\cite{levitt1971transformed}: 2 consecutive correct answers were required to increase the radius of the obstruction, while each incorrect answer would result in a decrease. This method allows for a 70.7\% performance accuracy measurement of threshold radius for each stimulus. Fig.~\ref{fig:fig_4} shows an example of the staircase for one of the participants. Each measurement was ended after either 14 staircase reversals (i.e., the radius of the obstruction changes from increasing to decreasing or vice versa) or 90 total trials. Initially, the obstruction was set to a radius of $\approx0.45\degree$. The step size of the change in the central obstruction radius in visual degrees was 30 pixels ($\approx1.35\degree$) up to the third reversal, then 20 pixels ($\approx0.9\degree$) up to the sixth, 10 pixels ($\approx0.45\degree$) up to the ninth, and 5 pixels ($\approx0.225\degree$) after nine reversals. Furthermore, the participant was randomly at a rate of 10\% shown a stimuli with an obstruction radius either 10 pixels ($\approx0.45\degree$) or 30 pixels ($\approx1.35\degree$), whose results were not considered. Note that the subjective retinal eccentricity to pixel size conversion values were determined using the structured light imaging method of Ref.~\cite{kapahi2024measuring}. The final threshold radius was calculated as the arithmetic mean of the final six reversal points. If the participant completed the maximum number of 90 trials, the final point was treated as a reversal point. 

Any participant whose results contained two or more reversals at the minimum obstruction radius within the final 6 reversals was assigned a ``failed'' status for that particular data point, which was removed from further analyses. Further, any data point lying outside a 99.9\% confidence interval ($\pm~3.3\sigma$) was excluded. After failures and outliers were removed, any participant with two or fewer retinal eccentricity threshold values remaining were also excluded because their results could not be reliably modelled. In total, 15 participants were included in the final analysis.% In total 5, 6, and 4 participants had 3, 4, and 5 valid threshold values respectively.

\section{Model}

\begin{figure}
    \centering
    \includegraphics[width=\linewidth]{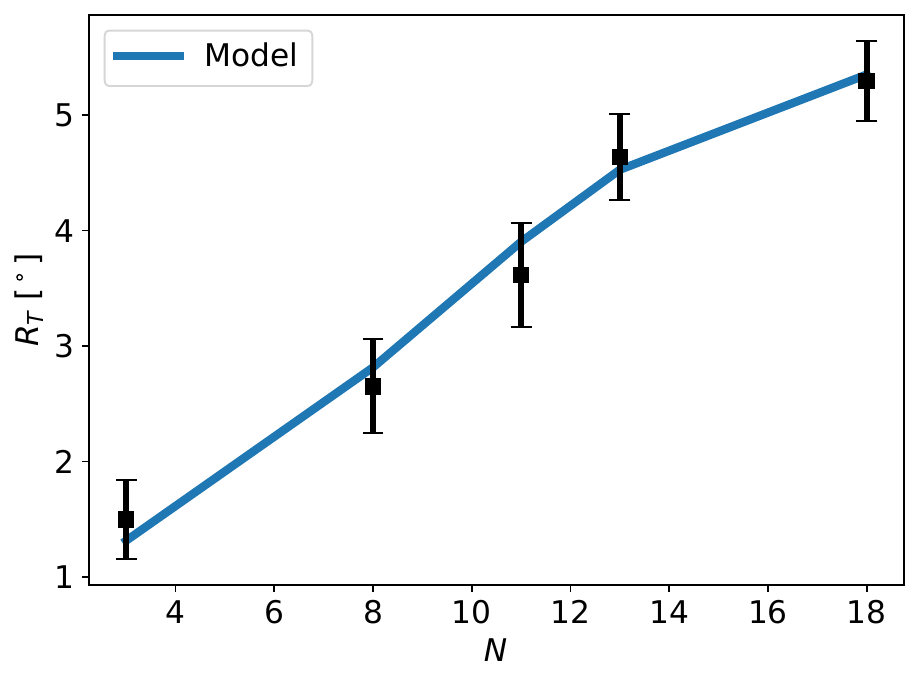}
    \caption{The average of the retinal eccentricity threshold results for the five different stimuli presented in this study: $(N=3,\ \omega=1.9~\text{Hz})$, $(N=8,\ \omega=3.8~\text{Hz})$, $(N=11,\ \omega=5~\text{Hz})$, $(N=13,\ \omega=5.8~\text{Hz})$, $(N=18,\ \omega=7.7~\text{Hz})$. Also shown are the averages of threshold radii obtained through fitting the coMPOD Model for individual participants. The Pearson $\chi^2$ fit statistic comparing the coMPOD model and the data is approximately 0.06, indicating a good fit. % on a $\chi^2$ distribution on 4 degrees of freedom, indicating a good fit. %The mean squared error (MSE) between coMPOD Model and the data is approximately 0.040.
    }
    \label{fig:fig_2}
\end{figure}

Fig.~\ref{fig:fig_1} outlines the developed model for reconstructing coMPOD profiles using retinal eccentricity thresholds for structured light stimuli. The target stimulus is a radially-symmetric polarization profile with $N$ fringes and rotating with temporal frequency $\omega$ in Hz, corresponding to an angular velocity $2\pi\omega/N$. The stimulus passes through a participant's cornea and macular pigment before perception. The cornea induces birefringence, and a portion of total macular pigment which is circularly-oriented acts as a weak radial polarizer. The perceived intensity is given by:

\begin{align}
    I(r,\phi, t)=&\frac{1}{4\pi\sigma_p^2}e^\frac{-r^2}{\sigma_p^2}\Big(4+\Theta[r-R]P(r)\big[\cos(N\phi + 2\pi\omega t) \nonumber \\
    &+ \cos((4+N)\phi + 2\pi\omega t) \nonumber \\
    &+ 2\cos\beta\sin{2\phi}\sin((2+N)\phi + 2\pi\omega t) - 2\big]\Big) \label{intensity_eqn}
\end{align}

\noindent where $R$ is the radius of the central obstruction, $\sigma_p$ is the beam width, ($r$, $\phi$) are retinal eccentricity coordinates, $t$ is time, $\Theta[r]$ is the Heaviside step function, $\beta$ is the corneal birefringence-induced phase shift, and $P(r)$ is the efficiency of the radial filter arising from coMP. For negligible corneal birefringence or in cases where corneal birefringence is compensated for, $\beta\approx 0$ and the equation for the perceived intensity simplifies to:

\begin{equation}
    I(r,\phi, t)=\frac{e^\frac{-r^2}{\sigma_p^2}}{2\pi\sigma_p^2}\Big(2+\Theta[r-R]P(r)\big[\cos(N\phi + 2\pi\omega t)-1\big] \Big) \label{intensity_eqn2}
\end{equation}

The efficiency of the radial polarizer in the human eye can be modelled by:

\begin{equation}
   P(r)=1-10^{-\coMP(r)}
\end{equation} 

\noindent where $\coMP(r)$ is the profile of the coMPOD. 

To determine if an individual can resolve the fringes of the stimulus, we first calculate the radially varying contrast $V(r)$. For a given radial fringe pattern ${f(\phi,t)=A+B \cos(k_\phi \phi - k_t t)}$, the contrast is time invariant and defined as $V={B}/{A}$.

The contrast sensitivity function describes the lowest perceivable contrast $V_T(k_r, k_t)$ for different spatial and temporal frequencies~\cite{van2001vision}: 

\begin{equation}
    \text{CSF}(k_r, k_t)=\frac{1}{V_T(k_r, k_t)}.
\end{equation}

\noindent $V_T(k_r, k_t)$ has been experimentally determined for linear fringe patterns presented to the center of the visual field. To account for the effects of cortical magnification on contrast sensitivity outside the fovea, we can transform the given on-axis contrast sensitivity function to an equivalent representation outside central vision through $M$-scaling~\cite{virsu1982temporal,rovamo1979estimation,daniel1961representation,strasburger2011peripheral}. To construct the $M$-scaled contrast sensitivity function, $\text{CSF}_\text{M}(r,k_r,k_t)$, any spatial frequency from the on-axis CSF is transformed according to $k\rightarrow k/s(r)$. Here $s(r)\approx (1+0.42r)^{-1}$ is dimensionless such that, defining $M_0$ to be the cortical magnification at $r=0$, $s(r)\cdot M_0$ describes the projection of the visual field onto the visual cortex in dimensions of millimeters per visual degree.

\begin{figure}
    \centering
    \includegraphics[width=\linewidth]{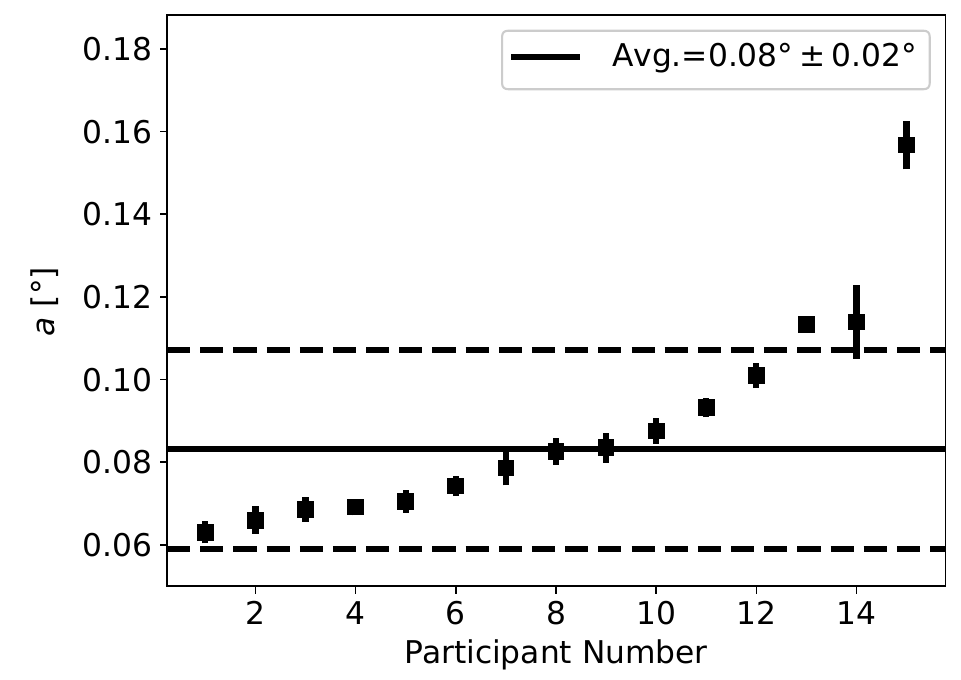}
    \caption{The obtained coMPOD parameter for each individual participant. The average $a$ was found to be $\overline{a}=0.08\degree\pm0.02\degree$. Shown in dashed lines are $\pm1$ standard deviation. Error bars on each data point denote an approximate 99.9\% confidence interval. 
    }
    \label{fig:fig_3}
\end{figure}

We can then define clarity, $C(r)$, to be the ratio of the stimulus contrast and the lowest perceivable contrast:

\begin{equation}
    C(r)\equiv V(r) \cdot \text{CSF}_\text{M}\left(r,\frac{N}{2\pi r}, \omega\right) \label{clarity_eqn}
\end{equation}

\noindent and the peak clarity of the stimulus:

\begin{equation}
    C_P\equiv \max[C(r)] \label{pattern_clarity_eqn}.
\end{equation} 

\noindent By definition, if $C_P<1$ the direction of the pattern's rotation is indiscernible. Since the efficiency of the eye's radially-polarized light filter decays quickly with $r$ in the region of interest, the pattern that an individual will perceive is sharpest near the center and contrast quickly decays as $r$ increases. Due to this decreasing nature of $C(r)$, the radius $r=R_T$ at which $C(R_T)=1$ defines the retinal eccentricity threshold of the particular stimulus.

\section{Results and Discussion} \label{discussion}

The presented technique is selectively sensitive to coMPOD, rather than total MPOD. The profile of the total MPOD has been studied elsewhere~\cite{berendschot2006macular}, and the analysis assuming a similar profile for the coMPOD is presented in the Appendix. These results and the geometry of the Henle fiber layer motivate modelling the coMPOD in the range of $1.5\degree$ to $5.5\degree$ retinal eccentricity by:

\begin{equation}
    \coMP(r)=a/(2\pi r),
    \label{eq:a_over_r}
\end{equation} 

\noindent where $a$ is a constant. Eq.~\ref{eq:a_over_r} assumes that the total number of Henle's fibers and the MP they contain does not vary on average in the region of interest; all that occurs is that these fibers spread out with eccentricity as more space becomes available. The $\coMP(r)$ in Eq.~\ref{eq:a_over_r} is therefore proportional to the density of fibers at radius $r$; the factor $a$ is influenced by the number of Henle's fibers as well as the efficiency of the radial polarization effect of the contained MP. Note that this form for the coMPOD is not suited for the low eccentricity region. coMPOD must be bound according to coMPOD~$\rightarrow 0$ as $r\rightarrow 0$. This implies some maximum optical density is reached near the center of the fovea before the decay at the very center.

%Eq.~\ref{eq:a_over_r} allows us to model the coMPOD in the region of interest with one parameter rather than two. The reduction in fit parameters allows for more robust fitting of participants with fewer data points.

Fig.~\ref{fig:fig_2} shows that the model can fit the average threshold values well; the Pearson $\chi^2$ fit statistic is approximately 0.06 with 4 degrees of freedom. Fig.~\ref{fig:fig_3} shows the individual participant fits for the parameter $a$ where the average $a$ was found to be $\overline{a}=0.08\degree\pm0.02\degree$. A comparison between this average profile for coMPOD and the average profile for total MPOD~\cite{berendschot2006macular} reveals that a small portion $(\approx 1/10)$ of total MP contributes to coMP for the low eccentricity values near $1.5\degree$, while at higher values of $\approx 5.5\degree$ they are roughly equal. 

\section{Conclusion}

We have demonstrated a novel method to selectively characterize the circularly-oriented macular pigment optical density (coMPOD), thereby providing a procedure to quantify the microstructure of the macula for applications in diagnosis of macular disease. Future work will consider a direct comparison to total MPOD measurements via auto-fluorescence. This would provide a general ratio between MPOD and coMPOD. Furthermore, while the probed retinal eccentricity region in this study was $1.5\degree$ to $5.5\degree$, future studies will also probe the coMPOD at lower eccentricities, particularly the $0\degree$ to $1.5\degree$ region where some distortions in vision are commonly reported~\cite{rowland2019ocular}. 

Early treatment of AMD and other macular diseases is crucial to minimizing or eliminating visual impairment. Therefore, a more selective method of early diagnosis would improve patient prognoses. The presented techniques can be applied to a wide range of stimuli and macular defects. Specific use-cases could include those of the early diagnoses of AMD~\cite{thomas2021age}, pathological myopia, manifesting as a tear in the macula~\cite{ueta2020pathologic}, and macular telangiectasia, which is often associated with changes in the macular pigment~\cite{issa2013macular}. 

\section*{Acknowledgements}

This work was supported by the Canadian Excellence Research Chairs (CERC) program, the Natural Sciences and Engineering Research Council of Canada (NSERC) grants [RGPIN$-2018-04989$], [RPIN$-05394$], [RGPAS$-477166$],  the Government of Canada’s New Frontiers in Research Fund (NFRF) [NFRFE$-2019-00446$], the Velux Stiftung Foundation [Grant 1188], the InnoHK initiative and the Hong Kong Special Administrative Region Government, and the Canada  First  Research  Excellence  Fund  (CFREF). 

\bibliography{OAM}

\onecolumngrid
\clearpage
\section*{Appendix}

\subsection{MPOD profile Model}

The efficiency of the radial polarizer in the human eye can be modelled by:

\begin{equation}
   P(r)=1-10^{-\coMP(r)}
\end{equation} 

\noindent where $\coMP(r)$ is the coMPOD. Here we assume that $\coMP(r)$ is described by an equation that has a similar  form as the total MPOD function $m(r)$~\cite{berendschot2006macular}: 

\begin{equation}
    m(r)=A_1 10^{-\rho_1 r} + A_2 10^{-\rho_2(r-\alpha_2)^2},
    \label{mpod1}
\end{equation}

\noindent where $A_1$ is the peak MPOD (at the fovea), $\rho_1$ describes the slope of the MPOD function, and $A_2$, $\rho_2$, and $\alpha_2$ describing the amplitude, slope, and location of the secondary ring. The average parameters for an aged population ($50\pm17$ years old) have been measured to be $A_1=0.31\pm0.12$, $\rho_1=0.38\invdeg\pm0.24\invdeg$, $A_2=0.11\pm0.08$, $\rho_2=1.2(\degree)^{-2}\pm1.1(\degree)^{-2}$, and $\alpha_2=0.70\degree\pm0.66\degree$~\cite{berendschot2006macular}. In the regime of $r$ this study investigated, the second term becomes negligible with typical parameters, so the MPOD function can be well approximated by:

\begin{equation}
    m(r)\approx A_1 10^{-\rho_1 r}
   \end{equation} 
   
\noindent Therefore, we take the coMPOD profile to be:

\begin{equation}
    \coMP(r)=A_1^* 10^{-\rho_1^* r}
    \label{eq:eq_appendix}
   \end{equation} 

\noindent where  $A_1^*$ is the peak coMPOD and $\rho_1^*$ describes the slope.

\begin{figure}
    \centering
    \includegraphics[width=.75\linewidth]{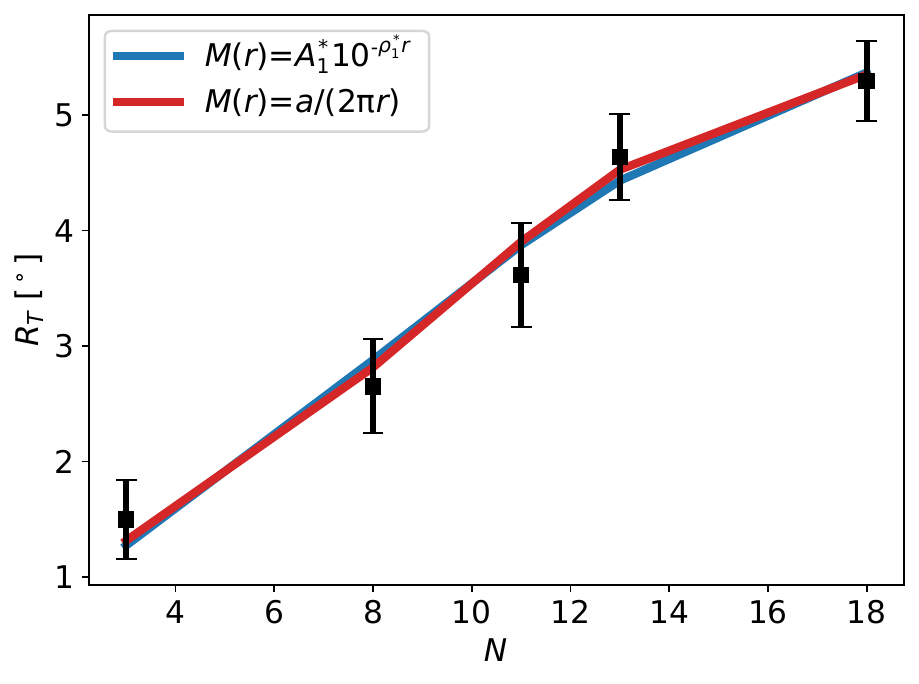}
    \caption{The average of the retinal eccentricity threshold results for the five different stimuli presented in this study: $(N=3,\ \omega=1.9~\text{Hz})$, $(N=8,\ \omega=3.8~\text{Hz})$, $(N=11,\ \omega=5~\text{Hz})$, $(N=13,\ \omega=5.8~\text{Hz})$, $(N=18,\ \omega=7.7~\text{Hz})$. Also shown are the averages of threshold radii obtained through fitting Eq.~\ref{eq:eq_appendix} (blue line) and Eq.~\ref{eq:a_over_r} (red line) for individual participants. The Pearson $\chi^2$ fit statistic for Eq.~\ref{eq:eq_appendix} is approximately 0.09 (with 3 degrees of freedom). That of Eq.~\ref{eq:a_over_r} is approximately 0.06 (with 4 degrees of freedom), indicating that both are good fits, though the Eq.~\ref{eq:a_over_r} fit is significantly more robust as it considers one fit parameter rather than two fit parameters.}%The mean squared error (MSE) between MPOD Model and the data is approximately 0.061.The MSE between coMPOD Model and the data is approximately 0.040, lower than that of the MPOD profile-like model.}
    \label{fig:fig_A2}
    
\end{figure}

\begin{figure}
    \centering
    \includegraphics[width=\linewidth]{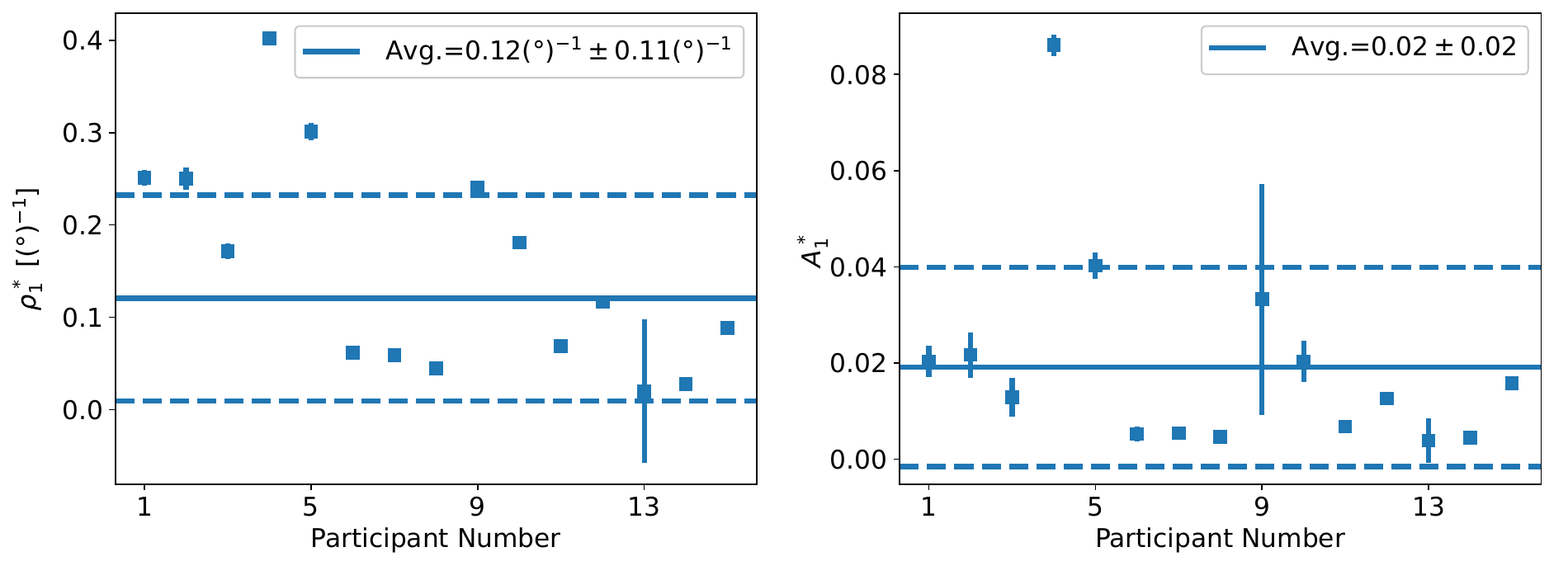}
    \caption{The obtained coMPOD parameters for each individual participant and the average using Eq.~\ref{eq:eq_appendix}. The average parameters were found to be $\overline{\rho_1^*}=0.12\invdeg\pm0.11\invdeg$ and $\overline{A}_1^*=0.02\pm0.02$. The average $A_1^*$ being significantly lower than the average value of $A_1$ (peak total MPOD) in Ref.~\cite{berendschot2006macular} implies that coMP is a small portion of total MP. Uncertainties in the fits are significantly more varied, as the absence of data points in some participants has a greater effect when fitting two parameters. Shown in dashed lines are $\pm1$ standard deviation. Data is ordered according to Fig.~\ref{fig:fig_3} in the manuscript. Error bars on each data point denote an approximate 99.9\% confidence interval. 
    }
    \label{fig:fig_A3}
   
\end{figure}

Fig.~\ref{fig:fig_A2} shows the average values of $R_T$ for study participants, as well as averages of $R_T$ fit to individuals' data using MPOD profile of Eq.~\ref{eq:eq_appendix}, for the free parameters $\rho_1^*$ and $A_1^*$. Comparing the two sets of values yields a Pearson goodness of fit statistic of $0.09$ on 3 degrees of freedom, showing good agreement between the data set and the model fits. Also shown on Fig.~\ref{fig:fig_A2} for comparison is the fit presented in the manuscript with $\coMP(r)=a/(2\pi r)$.

Fig.~\ref{fig:fig_A3} shows the fitted values of $\rho_1^*$ and $A_1^*$ for each participant. These values were used for modelling the average values of $R_T$ as shown in Fig.~\ref{fig:fig_A2}. We find that the average value of individuals' $\rho_1^*$ is ${0.12\invdeg\pm\ 0.11\invdeg}$, and the average of individuals' $A_1^*$ is approximately ${0.02\pm0.02}$, which is significantly lower than the peak MPOD found in Ref.~\cite{berendschot2006macular}. Noting the approximation $A 10^{-.14 r}\approx A/r$ for $r\in \{1.5,5.5\}$, motivates the use of the coMPOD form as described in the manuscript. The latter is significantly more robust as it considers one fit parameter rather than two fit parameters. %The lower value of $A_1^*$ as compared with the $A_1$ average found in Ref.~\cite{berendschot2006macular} implies that the portion of macular pigment contributing to the radially-polarizing effect is rather small.

\end{document}